\documentclass[fleqn,twoside,twocolumn,nofootinbib,showkeys]{revtex4} 
\usepackage[sec,nocpr]{ujp} 

\begin{document}
\title[Diffusion on the Distorted Fermi Surface]
{DIFFUSION ON THE DISTORTED FERMI SURFACE}
\author{V.M. Kolomietz}
\affiliation{Institute for Nuclear Research, Nat. Acad. of Sci. of
Ukraine}
\address{47, Prosp. Nauky, Kyiv 03680, Ukraine}
\email{lukyanov@kinr.kiev.ua}
\author{S.V. Lukyanov}
\affiliation{Institute for Nuclear Research, Nat. Acad. of Sci. of
Ukraine}
\address{47, Prosp. Nauky, Kyiv 03680, Ukraine}
\email{lukyanov@kinr.kiev.ua}

\udk{539.142}  \razd{\secii}

\autorcol{V.M.\hspace*{0.7mm}Kolomietz, S.V.\hspace*{0.7mm}Lukyanov}

\setcounter{page}{764}

\begin{abstract}
The diffusion approximation to the relaxation on the distorted Fermi
surface in a Fermi liquid is considered.\,\,The dependence of the
relaxation time on the multipolarity of a Fermi surface deformation
is established.\,\,The time evolution of the non-equilibrium
particle-hole excitations is studied.
\end{abstract}


\maketitle

\section{Introduction}\vspace*{1mm}

An important aspect of the dynamics of a Fermi liquid is the
presence of the Fermi motion of particles and the related effects of
the dynamic Fermi-surface distortions.\,\,Both can be considered in
the fluid dynamic approximation, where the initial quantum
mechanical equations of motion are converted into equations of
motion for local quantities such as the nucleon density, current
density, pressure,
\textit{etc}.\,\,\cite{bert75,kota81}.\,\,Usually, this is done by
means of the Fermi-surface deformation.\,\,The fluid dynamic
approximation gives a simple and clear-cut interpretation for the
highly collectivized giant resonances and establishes a relation
between the microscopic theory and phenomenological models such as
the the liquid-drop one \cite{echo81,kolo83,mako95,kiko96}.

The allowance of Fermi surface distortions leads to some features of
the Fermi-liquid dynamics, for example, the the excitation of transverse waves \cite%
{bert75,echo81,kolo83}, effect which do not occur in the non-viscous
classical hydrodynamics.\,\,Certain difficulties arise when one
tries to describe the damping of the collective motion, where the
macroscopic nuclear fluid dynamic description is far from being
established.\,\,The extension to the time-dependent mean-field
theory which incorporates interparticle collisions into the kinetic
equation for the many-body Fermi system leads to the necessity to
consider the dynamic Fermi-surface distortion in the collision
integral \cite{mako95,bert78,kopl96}.

In this paper, we used the diffusion approximation \cite{lipi81} to
the relaxation on the deformed Fermi surface.\,\,This approach gives
a simple result for the dependence of the relaxation time as a
function of the Fermi surface deformation multipolarity.\,\,In
Section 2, we consider the kinetic equation for the Wigner
distribution function and reduce it, by applying the diffuse
approximation.\,\,In Section 3, we establish the dependence of the
relaxation time on the Fermi-surface distortion multipolarity and
study the relaxation of particle-hole excitations.\,\,Our
conclusions are presented in Section 4.

\section{Diffuse Approximation\\ to the Collisional Kinetic Equation}

We will start from the collisional kinetic equation in the following form:
\cite{kaba62,abkh59}
\begin{equation}
\frac{\partial f_{1}(\mathbf{R},\mathbf{p}_{1};t)}{\partial t}+\hat{L}f_{1}
(\mathbf{R},\mathbf{p}_{1};t)=\delta \mathrm{St}\{f\}.
\label{lv}
\end{equation}
Here, $f(\mathbf{r},\mathbf{p};t)$ is the Wigner distribution
function, $\delta{\rm St}\{f\}$ is the collision integral, and the
operator $\hat{L}$ is given by
\begin{equation}
\hat{L}={\frac{1}{m}}\,\mathbf{p}\cdot \mathbf{\nabla }_{\mathbf{R}}
-(\mathbf{\nabla }_{\mathbf{R}}U)\cdot \mathbf{\nabla }_{\mathbf{p}}.
\label{2}
\end{equation}
The single-particle potential $U$ includes, in general,
self-consistent and external fields.\,\,We will use the collision
integral $\delta \mathrm{St}\{f\}$ in Eq.\,(\ref{lv}) in the
following general form:\vspace*{-1mm}
\[
\delta \mathrm{St}\{f\}=\int
{\frac{g^{2}d\mathbf{p}_{2}d\mathbf{p}_{3}d \mathbf{p}_{4}}{(2\pi
\hbar )^{6}}} w(\{\mathbf{p}_{j}\})Q(\{f_{j}\}) \,\times
\]\vspace*{-7mm}
\begin{equation}
\times\,\delta (\Delta \epsilon )\delta (\Delta \mathbf{p}).
\label{int1}
\end{equation}
Here, $w(\{\mathbf{p}_{j}\})\equiv w(\mathbf{p}_{1},\mathbf{p}_{2};\mathbf{p}%
_{3},\mathbf{p}_{4})$ is the spin-isospin averaged probability for the two-body
scattering, $g=$ $=4$ is the spin-isospin degeneracy factor, $%
Q(\{f_{j}\})=$
$=f_{1}f_{2}(1-f_{3})(1-f_{4})-(1-f_{1})(1-f_{2})f_{3}f_{4}$ is
the Pauli blocking factor, $\Delta \mathbf{p}=\mathbf{p}_{1}+\mathbf{p}_{2}-%
\mathbf{p}_{3}-\mathbf{p}_{4},$ and $\Delta \epsilon =\epsilon _{1}+\epsilon
_{2}-\epsilon _{3}-\epsilon _{4}$, with $\epsilon _{j}=p_{j}^{2}/2m+U(r_{j})$
being the single-particle energy.\,\,With the integration over the initial $%
\mathbf{p}_{2}$  and final $\mathbf{p}_{4}$ momenta of the medium
particles in the collision integral, one can reduce the kinetic equation (%
\ref{lv}) to the master equation with gain and loss terms.
Namely,\vspace*{-1mm}
\[
\frac{\partial f_{1}(\mathbf{R},\mathbf{p}_{1};t)}{\partial
t}+\hat{L}f_{1}(\mathbf{R},\mathbf{p}_{1};t) =\int
{\frac{\mathsf{g}d\mathbf{p}_{2}}{(2\pi \hbar )^{3}}} \,\times
\]\vspace*{-8mm}
\[ \times\, \bigl[W_{2\rightarrow 1}(\mathbf{p}_{1},\mathbf{p}_{2})
\tilde{f}_{1}(\mathbf{R},
\mathbf{p}_{1};t)\,f_{2}(\mathbf{R},\mathbf{p}_{2};t)\, -
\]\vspace*{-8mm}
\begin{equation}
-\, W_{1\rightarrow
2}(\mathbf{p}_{1},\mathbf{p}_{2})\,\tilde{f}_{2}\,(\mathbf{R},\mathbf{p}_{2};t)
f_{1}(\mathbf{R},\mathbf{p}_{1};t)\bigr]\!, \label{1}
\end{equation}
where $\tilde{f}_{j}=1-f_{j}$.\,\,The value $W_{i\leftrightarrows j}(\mathbf{p}%
_{1},\mathbf{p}_{2})$ in Eq.\,(\ref{1}) gives the transition
probability for
the scattering of a particle from the state $\mathbf{p}_{i}$ to the state $%
\mathbf{p}_{j}$ in surroundings of medium particles.\,\,The probability $%
W_{i\leftrightarrows j}(\mathbf{p}_{1},\mathbf{p}_{2})$ contains the
square of the corresponding amplitude of scattering for the direct,
\mbox{$1\rightarrow 2$}, and the reverse, \mbox{$2\rightarrow 1$},
transitions which are the same because
of the detailed balance.\,\,The probability $W_{i\leftrightarrows j}(\mathbf{p}%
_{1},\mathbf{p}_{2})$ includes also the distribution functions of
the scattered particle in the initial and final states.\,\,In the
lowest orders of a change of the distribution function, we can put
the equilibrium distribution function
$f_{\mathrm{eq}}(\mathbf{r},\mathbf{p}),$ which is isotropic in the momentum space, into $W_{i\leftrightarrows j}(%
\mathbf{p}_{1},\mathbf{p}_{2})$. We also assume that the main
contribution to the scattering amplitude is given by the transitions
that correspond to a small momentum transfer: \mbox{$|
\mathbf{p}_{1}-\mathbf{p}_{2}| \ll p_{\rm F}$}, where $p_{\rm F}$ is
the Fermi momentum, see also Ref.\,\,\cite{abkh59}.\,\,Finally,
assuming Born's approximation for the scattering amplitude, we can
write the expansion\vspace*{-1mm}
\[
W(\mathbf{p}_{1},\mathbf{p}_{2})=W(|\mathbf{p}_{1}+\mathbf{s}/2|,s)=W(p_{1},s)\,+
\]\vspace*{-7mm}
\[+\,{\frac{1}{2}}s_{\nu }\mathbf{\nabla }_{p_{1},\nu }\,W(p_{1},s)\,+
\]
\begin{equation}
+\,{\frac{1%
}{8}}s_{\nu }s_{\mu }\mathbf{\nabla }_{p_{1},\nu }\mathbf{\nabla }%
_{p_{1},\mu }W(p_{1},s)+... , \label{3}
\end{equation}%
where $\mathbf{s}=\mathbf{p}_{2}-\mathbf{p}_{1},$ and the summation with
respect to the repeated subscripts is understood.

Taking into account that the contribution to the collision integral
is given by states near the Fermi surface and keeping terms up to
the second order in the transferred momentum, we have
finally\vspace*{-1mm}
\[
{\frac{\partial f_{1}}{\partial t}}+\hat{L}f_{1} =-\mathbf{\nabla}
_{p_{1},\nu }\,\Bigl[ K_{p}\,(\mathbf{\nabla }_{p_{1},\nu
}\,\epsilon_{1})\, f_{1}\,\tilde{f}_{1}\,+
\]\vspace*{-7mm}
\begin{equation}
+\,f_{1}^{2}\,\mathbf{\nabla }_{p_{1},\nu }\,D_{p}\Bigr]+
\mathbf{\nabla }_{p_{1}}^{2}\,(f_{1}\,D_{p}) \label{4}
\end{equation}
from Eq.\,(\ref{1}). Here, the diffusion term $D_{p}$ in the
momentum space is defined as \cite{lipi81}\vspace*{-1mm}
\begin{equation}
D_{p}={\frac{1}{2}}\,\int d\mathbf{s}\,\,s^{2}\,W(p_{1},s).  \label{5}
\end{equation}%
The drift term $K_{p}$ in Eq.\,(\ref{4}) is connected to the diffusion term $%
D_{p}$\ by the relation\vspace*{-1mm}
\begin{equation}
K_{p}=\,{\frac{\partial }{\partial \epsilon _{1}}}\,D_{p}.  \label{6}
\end{equation}
The right-hand side of Eq.\,(\ref{4}) can be identified as a
collision integral\vspace*{-1mm}
\[
\mathrm{St}\{f\}=-\mathbf{\nabla}_{p_{1},\nu}\left[K_{p}\,(\mathbf{\nabla}_{p_{1},\nu}\epsilon_{1})
f_{1}\tilde{f}_{1}+f_{1}^{2}\mathbf{\nabla }_{p_{1},\nu
}D_{p}\right]+
\]\vspace*{-8mm}
\begin{equation}
+\,\mathbf{\nabla }_{p_{1}}^{2}\,(f_{1}\,D_{p}). \label{coll}
\end{equation}\vspace*{-5mm}

\section{Relaxation Time\\ for Distorted Fermi Surface}

The transport equation (\ref{4}) is nonlinear differential equation
for the Wigner distribution function
$f(\mathbf{r},\mathbf{p};t)$.\,\,We shall consider
this equation further in the limit $D_{p}=\mathrm{const,}$ $K_{p}=\mathrm{%
const},$ and rewrite it as\vspace*{-1mm}
\begin{equation}
{\frac{\partial f}{\partial t}}+\hat{L}\,f=-K_{p}\,\mathbf{\nabla }_{p,\nu
}\,f\,\tilde{f}\,\mathbf{\nabla }_{p,\nu }\,\epsilon +D_{p}\,\mathbf{\nabla }%
_{p}^{2}\,f.  \label{7}
\end{equation}

In the limit $\partial f/\partial t=0,$ this equation leads to the
correct equilibrium solution, which corresponds to a spherical Fermi
surface and gives the Fermi-type momentum distribution\vspace*{-1mm}
\begin{equation}
f=f_{\mathrm{eq}}(\mathbf{R},\mathbf{p}) =\left(\!
1+\mathrm{exp}{\frac{{p^{2}/2m+U-\lambda }}{T}}\!\right)
^{\!\!-1}\!\!, \label{8}
\end{equation}
where $\lambda $ is the chemical potential, which is derived by the
condition of the particle number $A$ conservation\vspace*{-1mm}
\begin{equation}
\int {\frac{\mathsf{g}d\mathbf{r}d\mathbf{p}}{(2\pi \hbar )^{3}}}f_{\mathrm{%
eq}}(\mathbf{r},\mathbf{p})=A.  \label{lambda}
\end{equation}%
This statement follows from definition (\ref{2}) of the ope\-rator
$\hat{L}, $
\begin{equation*}
\hat{L}\,f_{\mathrm{eq}}=0,
\end{equation*}%
and, further, from the fact that the right-hand side of Eq. (\ref{7})
\begin{equation}
\mathrm{St}\{f\}=-K_{p}\,\mathbf{\nabla }_{p,\nu }\,f\,\tilde{f}\,\mathbf{%
\nabla }_{p,\nu }\,\epsilon +D_{p}\,\mathbf{\nabla }_{p}^{2}\,f  \label{9}
\end{equation}%
for $D_{p}=\mathrm{const}$, $K_{p}=\mathrm{const}$ is equal to zero by $f=f_{%
\mathrm{eq}}$ and
\begin{equation}
T=-D_{p}/K_{p}.
\end{equation}
This last relation can be interpreted as a definition for the temperature $T$.

\begin{figure}
\vskip1mm
\includegraphics[width=\column]{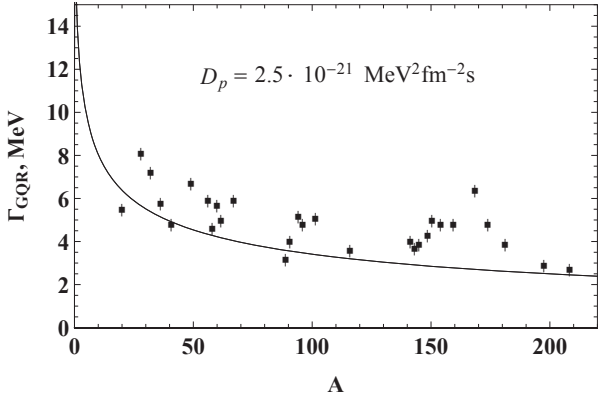}
\vskip-3mm\caption{Width of the isoscalar \textrm{GQR} versus the
nuclear mass number $A$.\,\,The results are obtained from
Eqs.\,\,(\ref{gamma1}) and (\protect\ref{tau1}) with $D_{p}(p_{\rm
F})=D_{p}=2.5\times10^{-21}$ $\mathrm{MeV}^{2}\, \mathrm{fm}^{-2}\,
\mathrm{s}$.\,\,The experimantal data are taken from Ref.
\cite{bert81}} \label{fig1}
\end{figure}

\subsection{Multipole deformation of Fermi surface}

In order to analyze the dependence of the collision integral $\mathrm{St}%
\{f\}$ and of the corresponding relaxation time on the multipolarity of a
Fermi surface distortion, we consider below a small deviation $\delta f=f-f_{%
\mathrm{eq}}$ of the distribution function from the
equilibrium.\,\,We shall expand this deviation in a series in
spherical harmonics such as
\begin{equation}
\delta f=-\frac{\partial f_{\mathrm{eq}}}{\partial \epsilon}
\,\sum_{lm}\,\nu_{lm}(\mathbf{r},t)\,Y_{lm}(\Omega_{p}).  \label{10}
\end{equation}
Putting this expanded form into the linearized collision integral
(\ref{9}), we obtain the multipole expansion of $\mathrm{St}\{f\},$
which contains all terms of $l$ starting from $l=0$.\,\,This comes
as a consequence of the above-used diffusion
approximation.\,\,However, the correct expression for the collision
integral does not contain the $l=0$ and $l=1$ terms because of the
conservations of the number of particles and the momentum in the
collisions.
With regard for these constraints, expansion (\ref{10}), and the expression for the collision integral (%
\ref{coll}) linearized with respect to $\delta f,$ we derive the relaxation time $\tau _{r,l}$ for the $l$-multipole
distortion of the Fermi surface:
\[
{\frac{1}{\tau _{r,l}}}=-\frac{\displaystyle \int
{\frac{d\mathbf{p}}{(2\pi \hbar )^{3}}}\,\,
\mathrm{St}\{f\}\,Y_{lm}(\Omega _{p})}{\displaystyle \int
{\frac{d\mathbf{p}}{(2\pi \hbar )^{3}}}\,\,\delta f\,Y_{lm}(\Omega
_{p})}=
\]\vspace*{-5mm}
\begin{equation}
={\frac{D_{p}(p_{\rm F})}{p_{\rm F}^{2}}}\,l\,(l+1), \quad l\geq 2.
\label{tau1}
\end{equation}
As seen from Eq.\,\,(\ref{tau1}), the drift term and, thereby, the
nuclear mean field $U$ do not contribute to the relaxation time
$\tau_{r,l},$ which is only derived by the diffusion on the
distorted Fermi surface.

One can also see from Eq.\,\,(\ref{tau1}) that the relaxation time
$\tau_{r,l}$ decreases, as the multipolarity $l$ of the
Fermi-surface deformation increases.\,\,The nuclear fluid dynamics
approximation (FLDA) for the isoscalar excitations corresponds to
the case where \mbox{$l=2$} is preferable, see
Ref.\,\,\cite{kosh04}.\,\,The fast damping of the higher
multipolarities of the Fermi surface distortion in
Eq.\,\,(\ref{tau1}) gives an argument for the applicability of the
FLDA to the description of highly collectivized nuclear excitations.

Relation (\ref{tau1}) can be used to adjust the diffusion coefficient $%
D_{p}(p_{\rm F})$ to the nuclear data.\,\,Within the FLDA, the
relaxation time $\tau _{r,l=2}$ derive the width $\Gamma
_{\mathrm{GQR}}$ of the isoscalar Giant Quadrupole Resonance (GQR),
see Ref.\,\,\cite{kosh04}.\,\,Namely,
\begin{equation}
\Gamma _{\mathrm{GQR}}=\frac{4\ \epsilon _{\rm F}\ \hbar
}{mr_{0}^{2}}\frac{\tau }{1+(\omega _{\rm R}\tau )^{2}}A^{-2/3},
\label{gamma1}
\end{equation}%
where $\epsilon _{\rm F}$ is the Fermi energy, $r_{0}$ is the mean
nucleon-nucleon distance in the nucleus, $\tau \equiv \tau
_{r,l=2,}$ and $\omega _{\rm R}$ is the GQR eigenfrequency.\,\,In
\figurename\ \ref{fig1}, we show the results of calculations and the
comparison with experimental data for the GQR width for the nuclei
through the Periodic table of elements.\,\,We have here adopted
$\epsilon _{\rm F}=40$~MeV, $r_{0}=1.2$ fm, and the experimental
value of the GQR energy $\hbar \omega _{\rm R}=63
A^{-1/3}$~MeV.\,\,The
relaxation time $\tau $ in Eq.\,(\ref{gamma1}) was taken from Eq.\,(\ref{tau1}%
) with $D_{p}(p_{\rm F})=D_{p}=2.5\times 10^{-21}$ $\mathrm{MeV}^{2}\,\mathrm{%
fm}^{-2}\,\mathrm{s}$.

As can be seen from Fig.~\ref{fig1}, the FLDA provides a quite
satisfactory description of the widths $\Gamma _{\mathrm{GQR}},$
where the quadrupole distortions of the Fermi surface are taken into
consideration, and the above-mentioned diffuse coefficient
$D_{p}(p_{\rm F})$ is used.

We point out a some peculiarity of the nuclear isovector
excitations.\,\,The
iso\-vec\-tor current is not con\-ser\-ved in the neutron-proton collisions \cite%
{anik83}, and the dipole distortion of the Fermi surface is
represented at the collision integral.\,\,Thus, the term $l=1$ gives
an additional
contribution to the relaxation of col\-lec\-ti\-ve iso\-vec\-tor excitations \cite%
{kota81,yadi89,diko99}.

\subsection{Particle-hole distortion of Fermi surface}

Another possibility for a distortion of the Fermi surface is the
initial non-equilibrium particle-hole excitation.\,\,We will
restrict ourselves by a nuclear matter, which is homogeneous in the
$\mathbf{r}$-space, and assume a spherical Fermi surface of radius
$p_{\rm F}$.\,\,The Fermi momentum is derived by the condition for
the particle number $A$ within a fixed volume $\mathcal{V}$
\[
\int\limits_{0}^{p_{\rm F}}\frac{4\pi \mathcal{V}\mathsf{g}}{(2\pi
\hbar )^{3}}\ p^{2}dp=A. \]%
The distorted particle distribution
$f_{\mathrm{in}}(p,t=0)$ for the particle-hole excitation at the
initial time $t=0$ is given by
\[
f_{\mathrm{in}}(p,t=0)=
\]\vspace*{-9mm}
\[
=\left[ 1-\theta (p-p_{1}^{\prime })+\theta(p-p_{2}^{\prime
})\right] \left[ 1-\theta (p-p_{\rm F})\right]+
\]\vspace*{-9mm}
\begin{equation}
+\left[ 1-\theta (p-p_{2})\right] \theta (p-p_{1})\theta (p-p_{\rm
F}), \label{fin}
\end{equation}
which means the particle located at $p_{1}<p<p_{2}$ and the hole excitation
at $p_{1}^{\prime }<p<p_{2}^{\prime }$ for the fixed $p_{1}>p_{\rm F}$ and $%
p_{2}^{\prime }<p_{\rm F}$, respectively.\,\,Note that the intervals
$\Delta p^{\prime }=p_{2}^{\prime }-p_{1}^{\prime }$ and $\Delta
p=p_{2}-p_{1}$ should be taken from the conditions
\[
\int\limits_{0}^{p_{\rm F}}\frac{4\pi \mathcal{V} \mathsf{g}
dp}{(2\pi \hbar )^{3}}p^{2}f_{\mathrm{in}}(p,t=0)=A-1,
\]\vspace*{-7mm}
\[
\int\limits_{p_{\rm F}}^{\infty }\frac{4\pi \mathcal{V}
\mathsf{g}dp}{(2\pi \hbar )^{3}}p^{2}f_{\mathrm{in}}(p,t=0)=1.
\]

\begin{figure}
\vskip1mm
\includegraphics[width=\column]{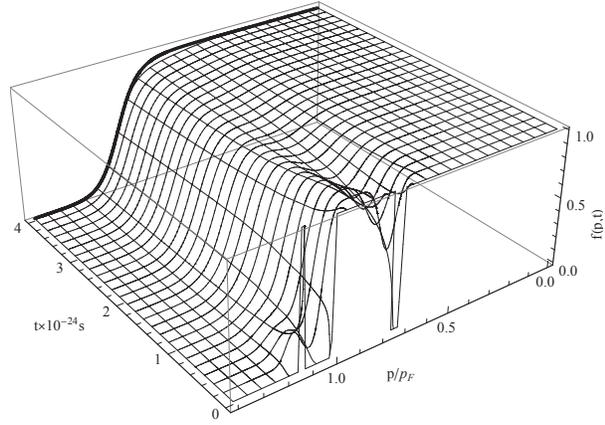}
\vskip-5mm\caption{Time evolution of the initial distribution
function (\ref{fin}) in the momentum space (in units of the Fermi
momentum $p_{\rm F}$) with $ p_{1}^{\prime }/p_{\rm F}\simeq 0.71$,
$p_{2}^{\prime }/p_{\rm F}\simeq 0.75$, $ p_{1}/p_{\rm F}\simeq
1.12$, \mbox{$p_{2}/p_{\rm F}\simeq 1.13$}, which corresponds to the
initial excitation energy $E_{\mathrm{ex}}=30$\,\,MeV.\,\,The solid
line is the Fermi distribution (\ref{8}) with $T=-D_{p}/K_{p}$. The
calculation was performed for $A=16$} \label{fig2}
\end{figure}

Assuming the spherical symmetric distribution, the kinetic equation (\ref{7}) is rewritten as
\begin{equation}
{\frac{\partial f}{\partial t}}=-\frac{K_{p}}{m}\,\left[ p\frac{\partial \,}
{\partial p}f\,\tilde{f}+3f\,\tilde{f}\ \right] +\frac{D_{p}}{p^{2}}\frac{%
\partial \,}{\partial p}p^{2}\frac{\partial \,}{\partial p}\,f.  \label{kin4}
\end{equation}%
Equation (\ref{kin4}) can be solved numerically under the initial
condition of Eq.\,(\ref{fin}).\,\,Below, we will use the transport
coefficients $D_{p}=2.5\times 10^{-21}\,\,\mathrm{MeV}^{2}\,\mathrm{fm}%
^{-2}\,\mathrm{s}$ and $K_{p}=-6.25\times
10^{-22}\,\,\mathrm{MeV}\,\mathrm{fm}^{-2}\,\mathrm{s}$.

In Fig.~\ref{fig2}, we have plotted the time evolution of the Wigner
distribution function $f(p,t)$ for the initial particle-hole
excitation in
a nucleus with $A=16$. Here, we have started from the initial distribution $%
f_{\mathrm{in}}(p,t=0)$ given by Eq.\,\,(\ref{fin}) and have assumed
the initial
excitation energy $E_{\mathrm{ex}}=30$ MeV.\,\,One can see from Fig.~%
\ref{fig2} that the momentum distribution $f(p,t)$ evolves to a
Fermi-type equilibrium limit $f_{\mathrm{eq}}(p)$ of the sort of
Eq.\,(\ref{8}).\,\,The corresponding equilibrium temperature of the
compound nucleus is obtained as $T=-D_{p}/K_{p}\approx 4\
\mathrm{MeV}$.

Note that the excitation energy $E_{\mathrm{ex}}$ is related to the
equilibrium temperature $T$ of the compound nucleus as $E_{\mathrm{ex}%
}=aT^{2}$, where $a$ is the statistical level density
parameter.\,\,In the case under study, we find $a\approx A/8.5$ MeV
$^{-1}$.\,\,The obtained value of the level density parameter $a$
agrees with the experimental one $a_{\exp }\simeq A/8$ MeV$^{-1}$
\cite{shko05} quite well.

\section{Conclusions}

We have considered the equilibration in a many-body Fermi-system
caused by the interparticle collision on the distorted Fermi
surface.\,\,Our approach combines both the dissipative and diffusive
effects providing the time evolution of the system toward the
equilibrium limit.\,\,We have discussed two types of non-equilibrium
states.\,\,The first one is the multipole deformation of the Fermi
surface, which occurs at the sound mode excitation.\,\,We have
established that the relaxation time $\tau _{r,l}$ decreases rapidly
with the growing multipolarity $l$ of the Fermi-surface distortion.
The relaxation time $\tau _{r,l}$ depends here on the diffuse
coefficient $D_{p}$ and does not depend on the drift coefficient
$K_{p}$.\,\,The second one is the relaxation of the particle-hole
excitation.\,\,We have shown that the
corresponding relaxation process depends on both transport coefficients $%
D_{p}$ and $K_{p},$ and the system drifts to the Fermi-type equilibrium limit
with the equilibrium temperature $T=-D_{p}/K_{p}$.

\vspace*{-5mm}
\rezume{%
В.М.\,Коломієць, С.В.\,Лук'янов}{ДИФУЗІЯ НА ДЕФОРМОВАНІЙ ПОВЕРХНІ
ФЕРМІ} {Розглянуто наближення дифузії для опису процесу  релаксації
на збуреній поверхні Фермі у фермі-рідині. Встановлено залежність
часу релаксації від мультипольності деформації поверхні Фермі.
Досліджено часову еволюцію нерівноважних збуджень частинка--дірка.}

\end{document}